\documentclass[aps,prb,twocolumn,superscriptaddress,showpacs,amsmath,amssymb, amsfonts,10pt]{revtex4-1}
\usepackage[T1]{fontenc}
\usepackage[utf8]{inputenc}
\usepackage{amsmath,bm}
\usepackage{amssymb}
\usepackage{yfonts}
\usepackage{eufrak}
\usepackage{bbm}
\usepackage{braket}
\usepackage{graphicx}
\usepackage{xcolor}
\usepackage{pifont}
\usepackage[mathscr]{euscript}
\usepackage[shortlabels]{enumitem}
\usepackage{float}
\usepackage{dsfont}
\usepackage{comment}
\usepackage{slashed}
\usepackage[normalem]{ulem} 
\usepackage{bbm} 

\usepackage[colorlinks=true]{hyperref}  
\hypersetup{
	unicode=false,          
	pdftoolbar=true,        
	pdfmenubar=true,        
	pdffitwindow=false,     
	pdfstartview={FitH},    
	pdftitle={FloquetAncilla},    
	pdfauthor={},     
	pdfsubject={},   
	pdfcreator={},   
	pdfproducer={}, 
	pdfkeywords={} {} {}, 
	pdfnewwindow=true,      
	colorlinks=true,       
	linkcolor=blue, 
	citecolor=blue,        
	filecolor=magenta,      
	urlcolor=blue           
} 

\usepackage{xcolor}
\definecolor{brandeisblue}{rgb}{0.0, 0.44, 1.0}
\usepackage{mathtools}

\newcommand{\bq}{\mathbf{q}}
\newcommand{\cl}{\mathrm{cl}}
\newcommand{\qn}{\mathrm{q}}
\newcommand{\rr}{\mathbf{r}}

\newcommand{\qq}{\mathbf{q}}
\newcommand{\bb}{\boldsymbol{B}}

\begin{document}
	\graphicspath{{figures/}}
	
	\title{Landau and fractionalized theories of periodically driven intertwined orders}
	
	\begin{abstract} 
		We obtain the phase diagrams of field theories of intertwined orders in the presence of periodic driving by an external field which preserves all symmetries. We consider both a conventional Landau theory of competing orders, and a fractionalized theory in which the order parameters are distinct composites of an underlying multi-component Higgs field.
		We work in the large $N$ limit and couple to a Markovian bath. The long time limits are characterized by non-zero average values, oscillations with the drive period and/or half the drive period, quasi-periodic oscillations, or chaotic behavior.
	\end{abstract}

	\author{Oriana K. Diessel}
	\affiliation{ITAMP, Center for Astrophysics | Harvard \& Smithsonian, Cambridge, MA 02138, USA}
	\affiliation{Department of Physics, Harvard University, Cambridge MA 02138, USA}
	
	\author{Subir Sachdev}
	\affiliation{Department of Physics, Harvard University, Cambridge MA 02138, USA}
	\affiliation{Center for Computational Quantum Physics, Flatiron Institute, 162 5th Avenue, New York, NY 10010, USA}
	
	\author{Pietro M. Bonetti}
	\affiliation{Max Planck Institute for Solid State Research, Heisenbergstraße 1, D-70569 Stuttgart, Germany}
	\affiliation{Department of Physics, Harvard University, Cambridge MA 02138, USA}
	
	\maketitle
	\newpage
	\linespread{1.05}
	\section{Introduction}
	\label{sec:intro}
	
	A number of experiments have studied the interplay between charge order and superconductivity in the presence of optical driving in the cuprate materials \cite{Cremin19,Shimano23,Shimano24}. Motivated by these results, the present paper extends our recent study \cite{Diessel25} of the periodic driving of a single order parameter with O($N$) symmetry to the case of two intertwined orders \cite{FradkinRMP}. 
	
	We will consider two models of the intertwined orders, a conventional Landau theory, and a fractionalized theory in which the orders are distinct gauge-invariant composites of an underlying Higgs field.  
	
	The conventional approach is based on a Landau theory framework, with two order parameters $\phi_1$ and $\phi_2$, and the interaction between them is controlled by a $\phi_1^2 \phi_2^2$. The periodic driving of such a theory is discussed in Section~\ref{sec:Landau}. This theory could apply in a variety of situations with multiple order parameters in numerous materials, including the pnictides and the electron-doped cuprates.
	
	The fractionalized approach in Section~\ref{sec:frac} is focused on the hole-doped cuprates. Following a proposal by Christos {\it et al.\/} \cite{Christos_2023,CS23,Christos2024,BCS24,ZhangSS24,Sayantan25,Boulder25}, both order parameters are represented by distinct gauge-invariant composites of a multi-component Higgs boson $\boldsymbol{B}$, which also transforms as a fundamental of an emergent SU(2) gauge field. If we include the full confining effects of the SU(2) gauge field, the effective theory for the order parameters reduces to that in the Landau theory of Section~\ref{sec:Landau}. However, our interest is in phenomena at distances shorter than the confining scale, where a fractionalized description is appropriate. Recent magnetotrasport experiments on the pseudogap metal in the hole-doped cuprates \cite{fang_admr_2022,chan_yamaji_2024} show strong evidence for small hole pockets with quasiparticles which can tunnel coherently between the square lattice layers. There is a natural explanation for such hole pockets in the fractionalized description \cite{RKK07,SSZaanen,Zhao_Yamaji_25,FuChun25}, but they are hard to understand in a formulation which works directly with the order parameters. So our analysis in Section~\ref{sec:frac} will carried out directly with the fractionalized Higgs field, and we will ignore the SU(2) gauge field, apart from a discussion of the electromagnetic response where it is required to obtain a properly gauge-invariant result. 
	
	Both computations here are carried out in a self-consistent large $N$ approach, similar to that in our previous paper \cite{Diessel25}. This approximation neglects fluctuations leading to long-time thermalization which could eventually modify some of the phases found. For the case of a single order parameter, the nature of the true long-time limit has been discussed \cite{Zelle2024,Daviet2024,Zelle25,MillisRubio26}, and it would be interesting to extend such results to the intertwined order parameter models discussed here. There can also be important effects on the ordered phases from Goldstone modes \cite{ZelleMillis25} which are not captured by our present large $N$ analysis.

	\section{Order competition and coexistence: Equilibrium}
	
	\subsection{Equilibrium fractionalized theory}
	
	We will describe periodic driving using a simplified field theory which captures the fractionalized approach to intertwined orders \cite{Christos_2023,CS23,Christos2024,BCS24,ZhangSS24,Sayantan25,Boulder25} in a large $N$ limit. 
	In this framework, we have $N$ bosonic fields $B_n=(B_{n,1},\,B_{n,2})$ that transform as doublets under the SU(2) gauge group. 
	
	In the large-$N$ the SU(2) gauge field decouples from the dynamics of the chargon field. For this reason, we will neglect it in the body of the paper. The SU(2) gauge field is needed for a proper computation of the electromagnetic response, and this is discussed in Appendix~\ref{sec:em}. In particular, it ensures that when the chargon field condenses, we get a Meissner effect only if we are in a superconducting phase. In the rest of the paper, without loss of generality, we will assume the field expectation values to take the form 
	\begin{equation}
		\langle B_n \rangle = \begin{cases}
			&\sqrt{N}b_+ (1,0)\text{ for } n=1\,,\\
			&\sqrt{N}b_- (0,1)\text{ for } n=2\,,\\
			& 0 \text{ otherwise}\,,
		\end{cases}
	\end{equation}
	with $b_\pm\in\mathbb{R}$.
	
	The intertwined orders are superconductivity (SC) and charge density wave (CDW), and these defined to be bilinears of $b_\pm$ \cite{Christos2024}
	\begin{equation}
		\label{eq:OP_definition}
		\phi_\text{CDW} = b_+^2 - b_-^2, \quad \phi_\text{SC} = 2 b_+b_-. 
	\end{equation}
	Our analysis will be performed on the fractionalized order parameter model described by the Lagrangian \cite{Christos2024}
	\begin{equation}\label{eq:Lagrangian}
		\begin{split}
			\mathcal{L} = \sum_{n=1}^N\Big[\frac{1}{2}|D_\mu B_n|^2&-\frac{r(t)}{2} |B_n|^2  - \frac{u}{4N}(|B_n|^2)^2\\
			&-\frac{v(t)}{4N}|B_n^\dagger\vec{\sigma}B_n|^2\Big],
		\end{split}
	\end{equation}
	where $D_\mu=\partial_\mu+i \vec{A}_\mu\cdot\vec{\sigma}$ is the covariant derivative and $\vec{\sigma}$ are the Pauli matrices. 
	We will see that in the Landau theory, discussed in Section~\ref{sec:LandauL}, the tuning between orders is present instead already at quadratic order, implying that the presence of nearly degenerate orders requires less fine-tuning in the fractionalized theory.
	\begin{figure}[t]
		\centering
		\includegraphics[width=1\linewidth]{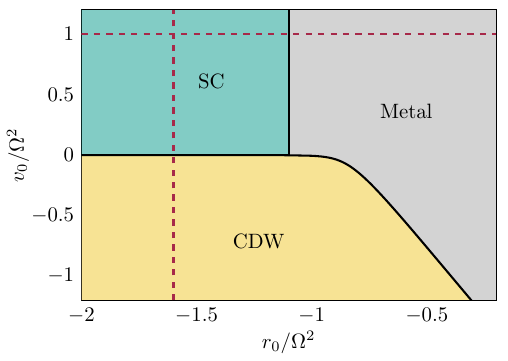}
		\caption{Equilibrium phase diagram of the fractionalized theory~\eqref{eq:Lagrangian} in the absence of gauge fields in the limit $N\to\infty$ \cite{Christos2024}. The red dashed lines indicate parameter cuts along which we study the effect of a periodic drive. The horizontal cut can be mapped to a theory with a single order parameter, corresponding to the driven $O(N)$ model analyzed in Ref.~\cite{Diessel25}. In this work we focus on the vertical cut, which probes the regime of two competing orders.}
		\label{fig:PD_Eq}
	\end{figure}
	The equilibrium phase diagram of this field theory at finite temperature can be calculated exactly in the $N \to \infty $ limit \cite{Christos2024} and is shown in Fig.~\ref{fig:PD_Eq}.
	As we are interested in two-dimensional materials, we focus on the model in two spatial dimensions. To circumvent the Mermin-Wagner theorem and allow for long-range order, we consider a system of finite size, which we implement by introducing an infra-red cutoff in our model.     
	
	Upon decreasing the parameter $r$, which in equilibrium controls the distance from the critical temperature, the systems undergoes a second-order phase transition from a disordered metallic phase for $r\geq r_c(v)$ to an ordered phase for $r <r_c(v)$, with $r_c(v)$ the $v$-dependent critical value of $r$. 
	For $v>0$, both $b_+$ and $b_-$ condense with the same value (up to a rotation), thus realizing a superconducting state where $\phi_\mathrm{SC}\neq0$, $\phi_\mathrm{CDW}=0$,
	while for $v<0$, only one of the field can condense, thus realizing a CDW phase ($\phi_\mathrm{SC}=0$, $\phi_\mathrm{CDW}\neq0$).. 
	The transition between the two ordered phases upon tuning the parameter $v$ is of the first order.
	
	We observe that this theory exhibits at equilibrium mostly competition between SC and CDW, and coexistence of these phases is only allowed on the critical line at $v=0$, where a first-order phase transition separates the two phases and coexistence is allowed.

	\subsection{Equilibrium O$(N)\times$O$(M)$-theory}
	\label{sec:LandauL}
	
	Turning to Landau theory approach to intertwined orders, the SC and CDW
	orders are now equal to the fields in the theory
	\begin{equation}
		\phi_{\rm CDW} = \phi_1 , \quad \phi_{\rm SC} = \phi_2\,.
		\label{eq:phiorders}
	\end{equation}
	We now have two real vector fields $\phi_1\in \mathbb{R}^N$ and $\phi_2\in \mathbb{R}^M$ 
	with a Landau–Ginzburg functional
	\begin{align}
		F=\frac{1}{2}(\nabla\phi_1)^2+\frac{1}{2}(\nabla\phi_2)^2+\frac{1}{2}r_1(t)\phi_1^2+\frac{1}{2}r_2(t)\phi_2^2\nonumber 
		\\+ \frac{u_1}{4N}\phi_1^4+ \frac{u_2}{4M}\phi_2^4+\frac{v(t)}{2\sqrt{NM}}\phi_1^2\phi_2^2.
		\label{Eq:Landau_model}
	\end{align}
	The quartic couplings are scaled with $1/N$ and $1/M$ in the standard way to ensure a well-defined large-$N,M$ saddle.
	
	This model has been extensively used to study the spin-flop transition in uniaxial ferromagnets~\cite{kosterlitz1976bicritical}, binary superfluids~\cite{timmermans1998phase},  and competing antiferromagnetism and superconductivity in correlated electrons~\cite{demler2004so}.
	The non-equilibrium, non-reciprocal version of this field theory has been discussed in Ref.~\cite{young2024nonequilibrium}.  
	
	\begin{figure}[h]
		\centering
		\includegraphics[width=1\linewidth]{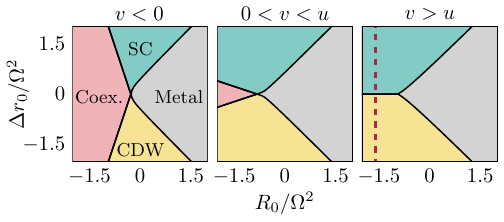}
		\caption{Equilibrium phase diagrams of the $O(N)\times O(M)$ model~\ref{Eq:Landau_model} as a function of the static mass parameters $R_0$ and $\Delta r_0$ for different inter-component couplings $v=-0.5$, $v=0.5$, and $v=1.5$. The red dashed line at $R_0/\Omega^2=-1.6$ corresponds to the parameters along which we study the effect of periodic driving.}
		\label{Fig:ONxON_eq}
	\end{figure}
	
	The model can be solved exactly in the large $N, M$ limit, and yields the phase diagram in Fig.~\ref{Fig:ONxON_eq}.
	For convenience, we reparametrized the bare masses in terms of a common and relative component,
	\begin{align}
		r_{1,2}= R\pm \Delta r,   
	\end{align}
	where $R$ controls the overall distance to criticality and $\Delta r$ controls relative detuning between the two order parameters. 
	We observe that the model presents coexistence, which is progressively taken over by competition upon increasing the value of $v$, until when no coexistence survives for $v>u$.  
	
	\subsection{Overarching questions}
	We want to adress two main questions:
	\begin{enumerate}
		\item Can a periodic drive suppress phase competition by stabilizing coexistence, or even by selecting a phase that would be unfavorable in equilibrium?
		\item What differences emerge in the phase diagrams of the two models under periodic driving? 
	\end{enumerate}
	
	\section{Driven fractionalized theory}
	
	\label{sec:frac}
	
	\subsection{Model and Keldysh action}
	We now consider a driven version of model~\eqref{eq:Lagrangian}. The couplings $r$ and $v$ become time dependent as 
	\begin{align}
		r(t) &=r_0+ 2r_1\cos(\Omega t) \nonumber \\
		v(t) &= v_0 + 2v_1\cos(\Omega t + \theta),
	\end{align}
	with $\Omega$ the driving frequency and $\theta$ the phase between the two driven parameters. Moreover, we couple the system to a classical bath at temperature $T$.
	
	Passing to the Keldysh path integral formalism~\cite{kamenev2023Book}, the classical action of the system, written in terms of the field expectation value $b_\pm$, is given by $S= \int_{\rr,t}(\mathcal{L}_0+ \mathcal{L}_\text{int})$, with $\mathcal{L}_0$ the Gaussian Lagrangian density given by (neglecting the SU(2) gauge field, which decouples at large $N$)
	\begin{equation}
		\begin{split}
			\mathcal{L}_0 \!=\! \sum_{\alpha =\pm}b_{\alpha,q}\big\{\big[-\partial_t^2 -\gamma\partial_t &+\nabla^2 - r(t)\big]b_{\alpha,c} \\ &+4i\gamma T b_{\alpha,q}\big\}\,,
		\end{split}
	\end{equation}
	where the terms proportional to $\gamma$ describe the coupling to the classical bath, and $\mathcal{L}_\text{int}$ the Gaussian Lagrangian density given by
	\begin{equation}
		\begin{split}
			\mathcal{L}_\text{int}= &-\sum_{\alpha = \pm}\frac{u+v(t)}{2N}(b_{\alpha,c}^2 + b_{\alpha,q}^2)b_{\alpha,c}b_{\alpha,q} \\
			&-\frac{u-v(t)}{2N}\left [b_{+,q}b_{+,c}(b^2_{-,c} \right. \\ & \left. +b^2_{-,q})+b_{-,q}b_{-,c}(b^2_{+,c}+b^2_{+,q})\right]\,.
		\end{split}
	\end{equation}
	Here, $b_{\alpha,c}$ and $b_{\alpha,q}$ refer to the classical and quantum components of $b_\pm$, respectively.
	\begin{figure}[h]
		\centering
		\includegraphics[width=1\linewidth]{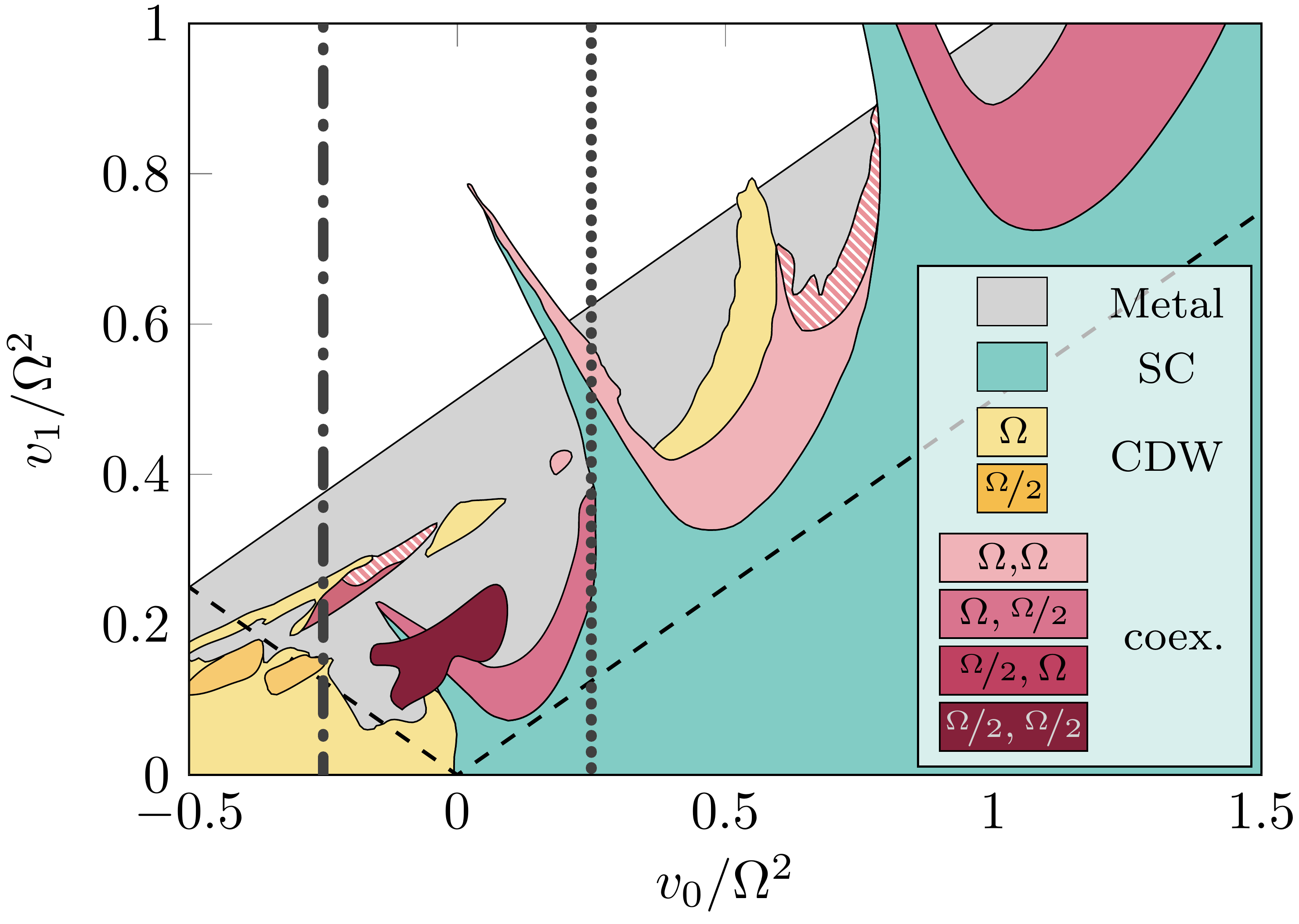}
		\caption{Phase diagram of the fractionalized theory in the $v_0/\Omega^2-v_1/\Omega^2$
			plane, obtained from solutions of Eqs.~\eqref{eq:EoM}. Different colors denote the metallic phase (gray), superconducting (SC) phase (blue), and charge-density-wave (CDW) phase (yellow). Regions with simultaneous SC and CDW order are shown in pink, with the labels indicating the dominant frequency components of the order parameters: $\Omega$ and/or $\Omega/2$. Hatched areas mark parameter regimes where neither simple $\Omega$-locked nor $\Omega/2$-locked responses occur, but instead signatures of quasiperiodic or chaotic dynamics are found (cf. Fig.~\ref{Fig:Chaos}). The dashed–dotted and dotted lines mark cuts along which the parameter $r_1/\Omega^2$	
			is additionally driven (see Fig.~\ref{Fig:PD_r1v1}).}
		\label{Fig:AncillaPD}
	\end{figure}
	
	\subsection{Large$-N$ limit and equations of motion}
	
	\begin{figure}
		\centering
		\includegraphics[width=1\linewidth]{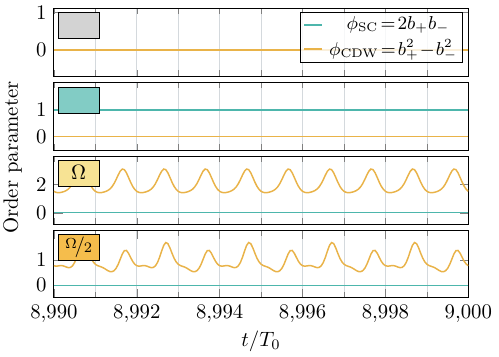}
		\caption{Time evolution of the order parameters during the final 10 drive periods of a simulation run over 9000 driving periods in the metallic (gray), SC (blue) and CDW (yellow) phases. From top to bottom, the panels correspond to  $(v_0, v_1)=(0.0, 0.4)$, $(0.5,0.2)$, $(-0.3,0.08)$, and $(-0.3,0.12)$.}
		\label{Fig:SingleOPAncilla}
	\end{figure}
	\begin{figure*}[t]
		\centering
		\includegraphics[width=1\linewidth]{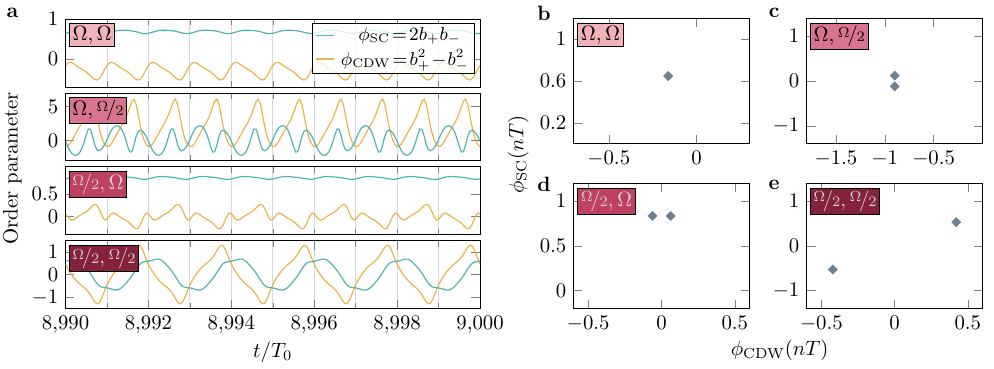}    
		\caption{Representative time evolution of the superconducting order parameter $\phi_{\text{SC}}$ (blue) and the charge-density-wave order parameter $\phi_{\text{CDW}}$ (yellow) in the four different coexisting phases.
			From top to bottom the panels show oscillations at $(\Omega,\Omega),(\Omega,\Omega/2),(\Omega/2,\Omega),$ and $(\Omega/2,\Omega/2)$, respectively. The corresponding parameter values are $(v_0,v_1) = (0.5,0.38), (-0.21,0.24), (1.2,0.8), (0,0.18)$.
			(b–e) Corresponding stroboscopic plots, where the order parameter is sampled at
			integer multiples of the drive period, $t=nT$, over the last $t/T=2000$ cycles.
			Each point corresponds to the pair $(\phi_{\text{CDW}}(nT),\,\phi_{\text{SC}}
			(nT))$. Panel (b) shows the phase where both order parameters oscillate at 
			$\Omega$. Panel (c) illustrates the coexistence of both components oscillating at
			different frequencies ($\Omega,\Omega/2$), while panel (d) shows the complementary
			mixed case with $(\Omega/2,\Omega)$ contributions. Panel (e) corresponds to the
			phase where both order parameters 
			are period-doubled ($\Omega/2,\Omega/2$).}
		\label{Fig:AncillaOP}
	\end{figure*}
	We now perform two Hubbard-Stratonovich transformations to decouple the interaction terms and then we derive the saddle-point equations in the large-$N$ limit. The derivation is shown in App.~\ref{app:HS_transformation}, and we report here only the final result.
	To this end, we assume a spatially homogeneous condensate and we posit $\langle b_{\pm,c} \rangle = (b_{\pm,c},,0,\dots,0)$. The equation of motion for the order parameters and for the correlation functions $C^\pm_\qq(t) = \langle b_\pm(\qq,t)^2\rangle$ and $D^\pm_\qq(t)=\langle \partial_t b_\pm(\qq,t)^2\rangle$ are given by:
	\begin{subequations}
		\label{eq:EoM}
		\begin{align}
			&\left[\partial_t^2+\gamma\partial_t + \lambda_\pm(t)\right]b_{\pm}(t)=0 \label{eq:EoM1}\\
			&\left[\partial_t^2+\gamma\partial_t+2(q^2 +\lambda_\pm(t))\right]C^\pm_\qq(t)=2D^\pm_\qq(t)\label{eq:EoM2}\\
			&\left(\partial_t+2\gamma\right)D^\pm_\qq(t)+\left[q^2+\lambda_\pm(t)\right]\partial_t C^\pm_\qq(t)=2\gamma T \label{eq:EoM3} 
		\end{align}
	\end{subequations}
	where $\lambda_\pm(t)$ is determined selfconsistently as:
	\begin{equation}
		\begin{split}
			\lambda_\pm(t) = r(t) &+ \left(\frac{u \pm v(t)}{2}\right)\left[b_+^2 + \int_\qq C^+_\qq(t) \right]\\
			&+  \left(\frac{u \mp v(t)}{2}\right)\left[b_-^2 + \int_\qq C^-_\qq(t)\right]\,,
		\end{split}
	\end{equation}
	where $\int_\bq$ is a shorthand for $\int_{\Lambda_\mathrm{IR}}^{\Lambda_\mathrm{UV}}\frac{d^2q}{(2\pi)^2}$.
	
	\subsection{Numerical results}
	
	We consider the vertical cut along the $v_0$ direction of the equilibrium diagram in Fig.~\ref{fig:PD_Eq}, corresponding to a transition from the CDW to the SC phase driven by the coupling $v_0$. We then assume it to be periodically driven as  $v(t) = v_0 + 2v_1\cos(\Omega t)$, while we keep the mass-like parameter $r$ undriven.  We correspondingly solve numerically the self-consistent equations of motions~\eqref{eq:EoM} and report the resulting phase diagram in the $v_0-v_1$ plane in Fig~\ref{Fig:AncillaPD}. In Fig.~\ref{Fig:PD_r1v1} we show an additional phase diagram obtained by driving \textit{both} the mass parameter $r$ and the interaction $v$.
	Our results show that the periodic drive give rise to a very rich phase diagram compared to the equilibrium one, as discussed in the following.
	
	\begin{figure*}[t]
		\centering    \includegraphics[width=1\linewidth]{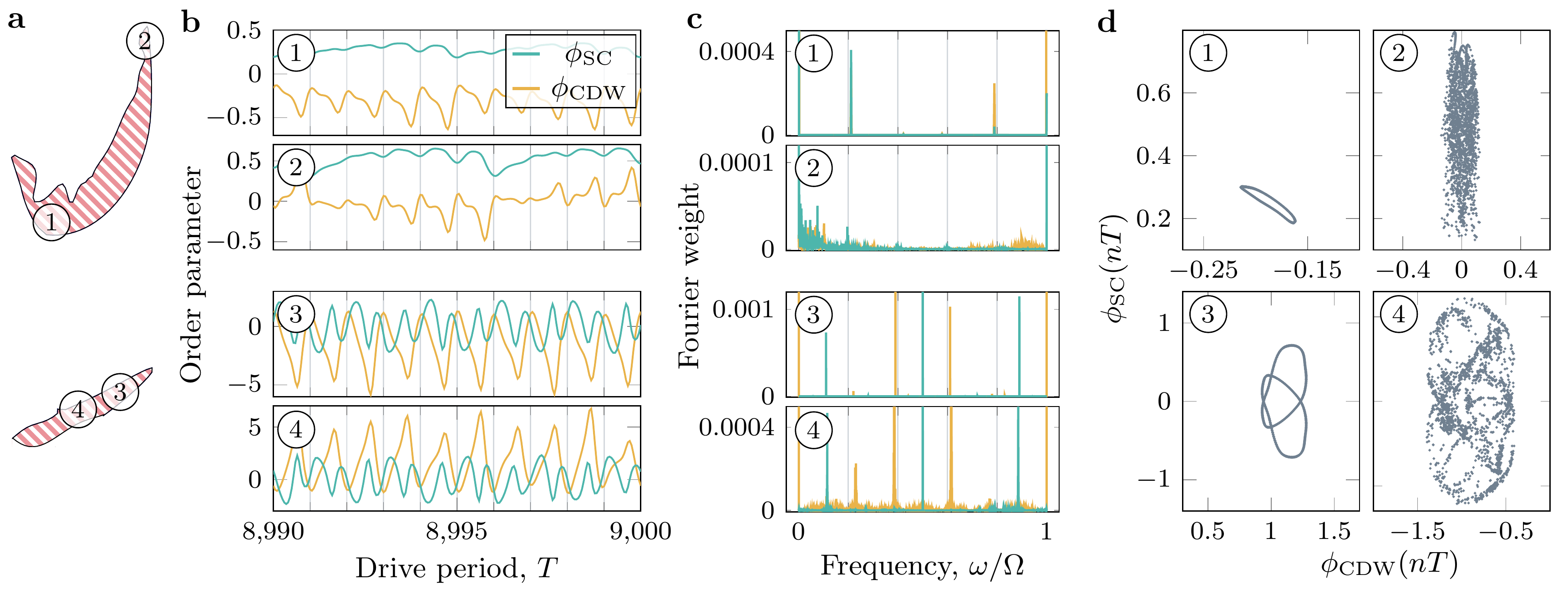}
		\caption{(a) Hatched regions of the phase diagram in Fig.~\ref{Fig:AncillaPD} indicating parameter domains where the dynamics do not settle into simple periodic phases. The numbered markers denote representative points with coordinates $(v_0,v_1)=(0.65,0.61),(0.78,0.87),(-0.08,0.31,)(-0.13,0.29)$, corresponding to points (1)-(4), respectively. (b) Time evolution of the order parameters $\phi_{\text{SC}}$ and $\phi_{\text{CDW}}$ over the last ten drive periods at the indicated points.
			(c) Corresponding Fourier spectra of $\phi_{\text{CDW}}$ and $\phi_{\text{SC}}$. At points (1) and (3), sharp peaks appear at frequencies incommensurate with the drive, characteristic of quasiperiodic dynamics. At points (2) and (4), the spectra display broad continua, indicating chaotic behavior.
			(d) Stroboscopic plots of $(\phi_{\text{CDW}}(nT),\phi_{\text{SC}}(nT))$ over the last 2000 drive periods. Quasiperiodicity is visible in panels (1) and (3) as trajectories tracing continuous closed curves, while chaotic attractors appear in panels (2) and (4) as diffuse clouds of points.}
		\label{Fig:Chaos}
	\end{figure*}
	\subsubsection{Arnold tongues} The first remarkable feature is the emergence of Arnold-tongue structures, which can be understood as footprints of the stability structure of the Mathieu-Hill equations, as discussed in Ref.~\cite{Diessel25}. We observe, moreover, that some tongue-structures appear ``nested'', {\it i.e.\/}, smaller tongue structures appear within larger tongue structures. 
	This can be understood as $b_+$ and $b_-$ obey different self-consistent Mathieu-Hill equations, which lead to two set of tongues structure of different size and with a finite offset.
	
	\subsubsection{Drive-induced suppression of competing orders}
	We observe that a strong enough periodic drive $v_1$ removes competition between SC and CDW orders and unlocks phases where SC and CDW coexist (region in different shades of pink).
	We found that this happens in the region where $v(t)$ oscillates between positive and negative values, i.e., where $v_0 + 2v_1>0$ and $v_0 - 2v_1<0$, corresponding to the region above the dashed lines. In the region below the dashed lines, $v(t)$ is always either completely positive (right side) or negative (left side), and competition between orders remains unsuppressed.  
	Additionally, we observe that the coexisting regions are located mostly close to the tips of the tongue structures.

	\subsubsection{Drive-induced metallic phase} The most internal part of the tongues shows the emergence of a metallic phase, which is not expected at equilibrium in this parameter regime.  
	
	\subsubsection{Time dependence of ordered phases} The SC phase (blue) does not exhibit any oscillations, as a vanishing CDW order parameter effectively decouples the model from the external drive (compare to Eq.~\eqref{eq:Lagrangian}). In the CDW phase, instead, we have both same-period and period-doubling solutions (different shades of yellow) (see Fig.~\ref{Fig:SingleOPAncilla}). Note that the period-doubling solution does not have a zero mean, in contrast to our previous work with a single order parameter \cite{Diessel25}. A nonvanishing offset in a period-doubling regime implies that period doubling occurs also in gauge-invariant quantities, such as the self-consistent mass $r(t)$, making the former potentially directly observable. 
	In the co-existing phase, we have four different combinations of SC and CDW oscillating with same-period or period doubling (different shades of pink) (see Fig.~\ref{Fig:AncillaOP}), as well small regions (hatched) where the dynamics is quasi-periodic or chaotic (see Fig.~\ref{Fig:Chaos}).
	
	\subsubsection{Unstable region}
	The upper-left corner of the phase diagram shows a region where no stable solution can be found. This easily understood as the region where the potential in the Lagrangian~\eqref{eq:Lagrangian} is unbounded from below and therefore unphysical. The equation of the line denoting the boundary of this region is given by $u+v_0 - 2|v_1| = 0$. However, we observe some of the re-entrant region of the tongues extends to this region, stabilizing the system.
	\begin{figure}[h]
		\centering
		\includegraphics[width=1\linewidth]{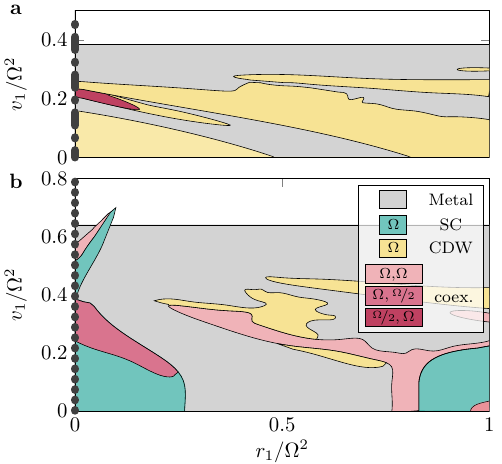}
		\caption{Phase diagrams with simultaneous driving of $r_1/\Omega^2$ and $v_1/\Omega^2$, for fixed $(r_0/\Omega^2,v_0/\Omega^2)=(-1.6,-0.25)$
			in \textbf{(a)} and $(r_0/\Omega^2,v_0/\Omega^2)=(-1.6,0.25)$ in \textbf{(b)}. Starting from a CDW equilibrium state at $(r_1/\Omega^2,v_1/\Omega^2)=(0,0)$, a pure SC phase cannot be reached \textbf{(a)}. In contrast, starting from an SC state, a pure CDW phase is accessible \textbf{(b)}. Thus, even driving both parameters does not remove the asymmetry observed Fig.~\ref{Fig:AncillaPD}.}
		\label{Fig:PD_r1v1}
	\end{figure}
	
	\subsubsection{Time-dependence of coexisting phases} In Fig.~\ref{Fig:AncillaOP}(a) we show representative time evolutions of the order parameters in each of the four coexisting  phases over the last 10 drive periods. The four phases display qualitatively distinct behavior: when the dynamics is locked to the drive frequency 
	$\Omega$, the order parameters oscillate about a finite mean value, whereas in the period-doubled phases the oscillations occur at half the drive frequency and are centered around zero.
	
	To make these distinctions more transparent, Figs.~\ref{Fig:AncillaOP}(b–e) present the corresponding stroboscopic plots, where the order parameter is sampled only at times 
	$t=nT$ over the last $2000$ drive periods (where the steady state is already reached). These plots provide a compact way of distinguishing the phases. A single point indicates a purely
	$\Omega$-locked response [panel (b)], while the appearance of two points indicates that at least one component oscillates with period 
	$2T$. The arrangement of the points indicates which component of the order parameter carries the period doubling. 
	In panel (e), both $\phi_{\text{CDW}}$ and $\phi_{\text{SC}}$  oscillate with period $2T$, yielding two points placed symmetrically about the origin. In panel (c), only $\phi_{\text{SC}}$ undergoes period doubling, while $\phi_{\text{CDW}}$ remains locked to the drive. Consequently,  the two stroboscopic points have identical values of $\phi_{\text{CDW}}$ but opposite values of $\phi_{\text{SC}}$. In panel (d), the situation is reversed: $\phi_{\text{CDW}}$ exhibits period doubling while $\phi_{\text{SC}}$ remains locked to the drive, producing two points with the same value of 
	$\phi_{\text{SC}}$ but different values of $\phi_{\text{CDW}}$.
	
	\subsubsection{Quasiperiodicity and chaos in the coexisting phases} 
	So far, we have identified phases characterized by well-defined periodic responses of the order parameters. In the hatched regions of the phase diagram~\ref{Fig:AncillaPD}, however, we do not encounter any of the above scenarios. Instead, the dynamics exhibit signatures of quasiperiodicity or even chaos.

	Representative examples are shown in Fig.~\ref{Fig:Chaos} (b–d). The time evolutions of the order parameters [panel (b)] exhibit irregular oscillations without a simple period, which is directly reflected in the Fourier spectra [panel (c)]. 
	At points (1) and (3), spectra feature sharp peaks at frequencies incommensurate with the drive frequency $\Omega$. This indicates quasiperiodic behaviour, which is also evident from the corresponding stroboscopic plots shown in panel (d). Instead of collapsing onto a discrete set of points, the trajectories trace out smooth, continuous curves.
	
	In contrast, at points (2) and (4), the Fourier spectra [panel (c)] display a broad continuum of frequencies, corresponding to chaotic dynamics. This manifests in the stroboscopic plots [panel (d)] as diffuse clouds of points forming irregular structures, which can be interpreted as chaotic attractors.
	
	\section{Driven $O(N)\times O(M)$ theory}
	\label{sec:Landau}

	We now consider the driven version of the equilibrium Landau $O(N)\times O(M)$ model described in Sec.~\ref{sec:LandauL}.
	In the limit $N,M\rightarrow\infty$, but $N/M=1$, the fields $\phi_i$ and correlation functions $C_i$ and $D_i$ obey the equations of motion
	\begin{subequations}
		\label{eq:EOM}
		\begin{align}
			&[\partial_t^2 +\gamma_i +\lambda_i(t)]\phi_i(t)=0\\
			&[\partial_t^2+\gamma_i+2(q^2+\lambda_i(t))]C_{i,\qq}(t)=2D_{i,\qq}(t)\\
			&(\partial_t+2\gamma_i) D_{i,\qq}(t)+[q^2+\lambda_i(t)]\partial_t C_{i,\qq}(t)=2\gamma_i T.
		\end{align}
	\end{subequations}
	The time-dependent effective masses $\lambda_i(t)$ are given by
	\begin{align}
		\lambda_1(t)=r_1(t)+u\left(\phi_1^2+\!\int_\qq\! C_{1,\qq}(t)\!\right)+v\!\left(\phi_2^2+\!\int_\qq\! C_{2,\qq}(t)\!\right)\\
		\lambda_2(t)=r_2(t)+u\left(\phi_2^2+\!\int_\qq\! C_{2,\qq}(t)\!\right)+v\!\left(\phi_1^2+\!\int_\qq\! C_{1,\qq}(t)\!\right).
	\end{align}
	
	These masses incorporate both mean-field contributions $\phi_i^2$ and fluctuations $\int_q C_i$, and they couple the two order parameters when $v\neq 0$.
	Similarly to the equilibrium case, it is convenient to reparametrize the bare masses in terms of a common relative component,
	\begin{align}
		r_{1,2}(t)= R\pm \Delta r(t).   
	\end{align}
	periodically driven as:
	\begin{equation}
		\Delta r(t)=\Delta r_0+2 \Delta r_1\text{cos}(\Omega t).   
	\end{equation}
	The resulting dynamics is obtained by solving Eqs.~\eqref{eq:EoM} in real time.
	To characterize the steady-state behavior, we extract order parameters from the late-time dynamics and construct nonequilibrium phase diagrams as a function of the static and driven relative mass. 
	
	\begin{figure}
		\centering
		\includegraphics[width=1\linewidth]{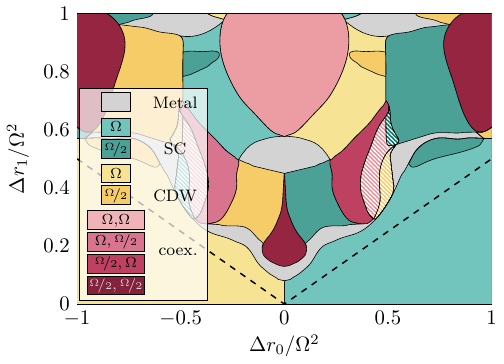}
		\caption{Phase diagram of the driven $O(N)\times O(M)$ model in the large $N,M$ limit, as a function of the static and driven mass components $r_{1,0}/\Omega^2$ and $r_{1,1}/\Omega^2$. Shown are the metal (gray), superconducting (SC, blue), charge-density-wave (CDW, yellow), and coexistence (coex., pink) phases. Parameters: $v=1.5$, $u_1=u_2=1$, bath temperature $T=1$, damping $\gamma_1=\gamma_2=0.2$, and drive frequency $\Omega=1$. The second component is driven symmetrically with 
			$r_{2,0}=r_{1,0}$, $r_{2,1}=-r_{1,1}$. Hatched regions indicate parameter regimes with nontrivial order parameter dynamics.}
		\label{fig:phase_diagram_O(N)xO(M)}
	\end{figure}

	\begin{figure}[t]
		\centering
		\includegraphics[width=1\linewidth]{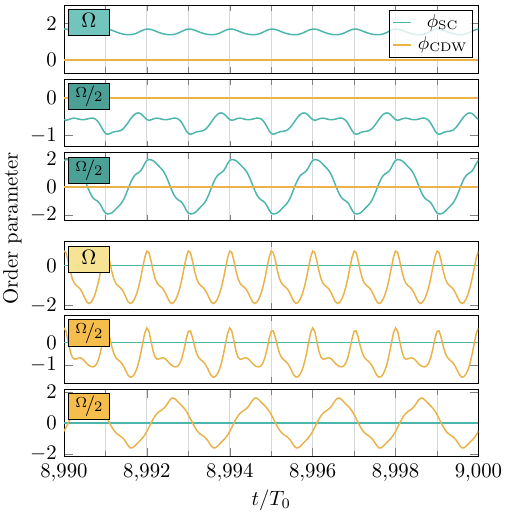}
		\caption{Time evolution of the order parameters during the final 10 drive periods of a simulation run over 9000 driving periods in the SC (blue) and CDW (yellow) phases. Period-doubled oscillations can occur around either zero mean or a finite offset in both phases. From top to bottom, the panels correspond to  $(\Delta r_0, \Delta r_1)=(0.5, 0.1)$, $(0.7,0.52)$, $(0.7,0.7)$, $(0.3,0.7)$, $(0.46,0.84)$ and $(-0.2,0.3)$, respectively.}
		\label{Fig:TimeEvolutionLandau}
	\end{figure}

	\label{sec:conclusions}
	
	\subsection{Numerical results}
	We consider the vertical cut along the $\Delta r_0$ direction of the equilibrium diagram in Fig.~\ref{Fig:ONxON_eq}, panel (c), corresponding to a transition from the CDW to the SC phase driven by $\Delta r_0$. We then assume it to be periodically driven as  $\Delta r(t) = \Delta r_0 + 2\Delta r_1\cos(\Omega t)$, while we keep the parameter $R$ undriven.  We correspondingly solve numerically the self-consistent equations of motion~\eqref{eq:EOM} and report the resulting phase diagram in the $\Delta r_0-\Delta r_1$ plane in Fig~\ref{fig:phase_diagram_O(N)xO(M)}. 
	Our results show that the periodic drive give rise to a very rich phase diagram compared to the equilibrium one, as discussed in the following.
	
	\subsubsection{Fine-tuned symmetry} For the parameters chosen, the phase diagram is symmetric under under $\phi_1 \leftrightarrow \phi_2, \Delta r_0 \to -\Delta r_0$. This is however a special fine-tuned case, and in general this symmetry is absent for a generic choice of the system parameters. Nevertheless, we work at this point in order to keep the Landau theory as similar as possible to the fractionalized field theory.

	\subsubsection{Drive-induced suppression of competing orders}
	We observe that a strong enough periodic drive $v_1$ removes competition between SC and CDW orders and unlocks phases where SC and CDW coexist (region in different shades of pink).
	We found that this happens in the region where $\Delta r(t)$ oscillates between positive and negative values, i.e., where $\Delta r_0 + 2\Delta r_1>0$ and $\Delta r_0 - 2\Delta r_1<0$, corresponding to the region above the dashed lines. In the region below the dashed lines, $\Delta r(t)$ is always either completely positive (right side) or negative (left side), and competition between orders remains unsuppressed.  
	
	\subsubsection{Drive-induced metallic phase} Some small regions shows the emergence of a metallic phase, which is not expected at equilibrium in this parameter regime. 

	\subsubsection{Time dependence of ordered phases} Both SC and CDW phase can have same-period and period-doubling solutions (different shades of blue and yellow), see Fig.~\ref{Fig:TimeEvolutionLandau}. 
	Period-doubled oscillations can occur around either zero mean or a finite offset in both phases. This contrasts with the single–order-parameter theory, where Floquet symmetry restricts period doubling to occur only around zero mean. In that case, the order parameter can be written in its Floquet representation
	$\phi(t) = e^{-i\nu t} \sum_{n} \phi_n e^{-i n \Omega t}$
	which enforces that period-doubled solutions appear only through harmonics around a vanishing average.
	In the presence of multiple coupled order parameters, however, the dynamics remain coupled through fluctuations even when one order parameter vanishes. This can induce period doubling of the effective mass and thereby allow the order parameter to oscillate with period 
	$2T$ around a finite value.  In the co-existing phase, we have four different combinations of SC and CDW oscillating with same-period or period doubling (different shades of pink), as well small regions (hatched) where the dynamics is quasi-periodic or chaotic.
	
	\begin{figure}
		\centering
		\includegraphics[width=1\linewidth]{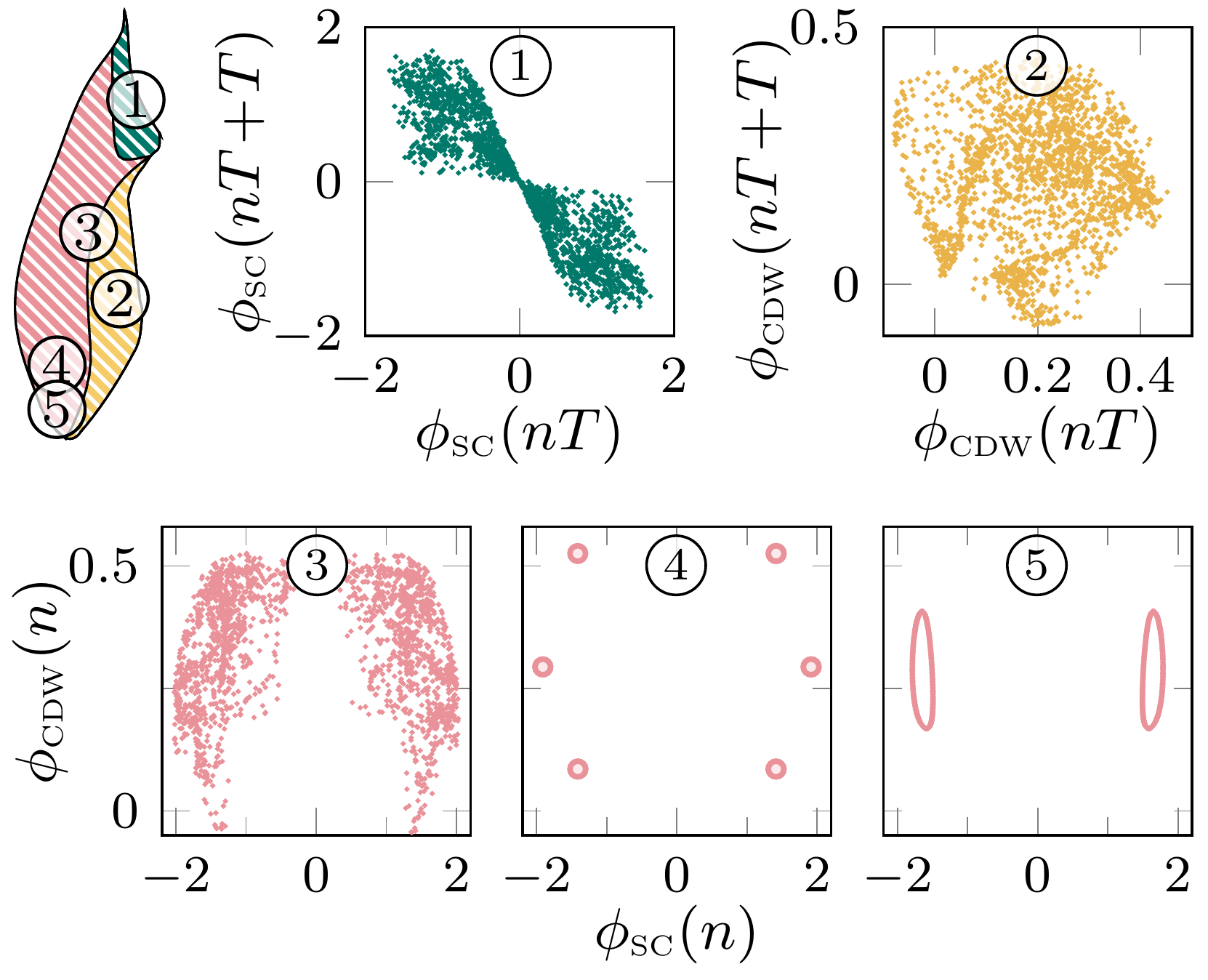}
		\caption{Hatched regions of the phase diagram~\ref{fig:phase_diagram_O(N)xO(M)} indicating parameter domains where the dynamics do not settle into simple periodic phases. The numbered points mark representative examples. To each numbered point the corresponding  stroboscopic plots of $(\phi_1(T),\phi_1((n+1)T))$ or $(\phi_1(nT),\phi_2(nT))$ are shown, displaying data over the last 2000 drive periods. Chaos is visible in (1)-(3), while quasiperiodicity is visible in (5). In (4), the stroboscopic plot shows six distinct points, indicating period-6 oscillations of the order parameters.} 
		\label{Fig:Stroboscopic plot}
	\end{figure}
	
	\subsubsection{Quasiperiodicity and chaos} In the hatched region in Fig.~\ref{fig:phase_diagram_O(N)xO(M)} the order parameters dynamics is characterized by quasiperiodic and chaotic behaviour. To further clarify this, we show in Fig.~\ref{Fig:Stroboscopic plot} the stroboscopic visualization of the order paramaters in that region. We first notice that the purely SC and CDW regions (blue and yellow, respectively) only exhibit chaotic behaviour, whose signature is a scattered collection of points in the stroboscopic plot. More interestingly, the coexistence phase (pink) show both chaotic and quasiperiodic behaviours, the latter being associated with closed lines of points in the corresponding stroboscopic plot.    
	
	\section{Discussion}
	
	We have presented self-consistent Floquet theories of intertwined order parameters, generalizing our earlier work on a single order parameter \cite{Diessel25}. Section~\ref{sec:frac} described the fractionalized theory in Eq.~(\ref{eq:Lagrangian}), where quadratic powers of the fundamental field $b_\alpha$ are related to the superconducting and charge density wave order parameters via Eq.~(\ref{eq:OP_definition}). Section~\ref{sec:Landau} described the conventional Landau theory in 
	Eq.~(\ref{Eq:Landau_model}), where the fundamental fields $\phi_{1,2}$ are now equal to the order parameters, as in Eq.~(\ref{eq:phiorders}).
	
	Throughout this work, we have restricted ourselves to spatially homogeneous Ansätze. Consequently, spatially inhomogeneous condensates are not captured within our approach.
	
	The phase diagram of the fractionalized case is in Fig.~\ref{Fig:AncillaPD} and \ref{Fig:PD_r1v1}, and that of the Landau case in Figs.~\ref{fig:phase_diagram_O(N)xO(M)}. Both cases exhibit the same set of phases, with oscillations of the two order parameters at one or both of the frequencies $\Omega/2$ and $\Omega$. Also notable are the hatched regions where is quasiperiodic or chaotic time evolution: such regions were not found in the single order parameter case \cite{Diessel25}.
	
	Ultimately the differences between the fractionalized and Landau cases reflect differences already present in the equilibiurm cases. Fig.~\ref{fig:phase_diagram_O(N)xO(M)} of the Landau theory is symmetric between the SC and CDW orders, while Fig.~\ref{Fig:AncillaPD} of the fractionalized has no such symmetry. The absence of a clear distinction between the two cases indicates that it is not unreasonable to use the Landau theory approach if only order parameters are being measured. But, as we noted in Section~\ref{sec:intro}, the fermionic spectrum, and associated transport properties, of the hole-doped cuprates are much more naturally described starting from the fractionalized theory \cite{RKK07,SSZaanen,Zhao_Yamaji_25,FuChun25}. The experiments of Refs.~\cite{Cremin19,Shimano23,Shimano24} show SC-like behavior induced by light on a CDW state: while our Landau theory allows for such behavior, the fractionalized phase diagrams in Figs.~\ref{fig:phase_diagram_O(N)xO(M)} and \ref{Fig:AncillaPD} show only small regions of co-existence starting from the CDW state. This suggests that ignoring the SU(2) gauge field may not be appropriate, and inclusion of fluctuations of the SU(2) gauge fields, as in Ref.~\cite{Sayantan25}, is needed.
	
	\subsection*{Acknowledgments}
	We gratefully acknowledge discussions with Alessio Chiocchetta and Carl P.~Zelle. This research was supported by NSF Grant DMR-2245246 and by the Simons Collaboration on Ultra-Quantum Matter which is a grant from the Simons Foundation (651440, S. S.). 
	The Flatiron Institute is a division of the Simons Foundation.
	O.K.D acknowledges support from the NSF through a grant for ITAMP at Harvard University.
	P.M.B. acknowledges support by the German National Academy of Sciences Leopoldina through Grant No.~LPDS 2023-06 and the Gordon and Betty Moore Foundation’s EPiQS Initiative Grant GBMF8683.
	
	\bibliography{biblio}
		
	\appendix
	
	\newpage
	\section{Hubbard-Stratonovich transformation and saddle point}
	\label{app:HS_transformation}
	We now perform two Hubbard-Stratonovich transformations to decouple the interaction terms to get the Lagrangian density

	\begin{subequations}
		\begin{align}
			\mathcal{L}_0=&(D_\mu\boldsymbol{B})^\dagger\kappa^1\,(D^\mu\boldsymbol{B})\nonumber\\
			&-\gamma \boldsymbol{B}^\dagger (-i\kappa^2)D_t \boldsymbol{B}+4i\gamma T |B_\qn|^2\\
			\mathcal{L}_\mathrm{HS}=&\mathcal{L}_0 
			+\frac{4N}{u}\lambda_q\lambda_c 
			+ \frac{4N}{v(t)}\vec{H}_q\vec{H}_c\nonumber
			\\&- \lambda_\cl 
			\left(|\bb_c|^2+|\bb_q|^2\right)\nonumber\\
			&-\lambda_\qn\left(\bb^\dagger_q\bb_c + \text{c.c.}\right)
			\\ &- \vec{H}_\cl\cdot\left(\boldsymbol{B}^\dagger_q\vec{\sigma}\boldsymbol{B} + \text{c.c.}\right) \nonumber\\
			&-\vec{H}_\qn\cdot\left(\boldsymbol{B}^\dagger_c\vec{\sigma}\boldsymbol{B}_c+\boldsymbol{B}^\dagger_q\vec{\sigma}\boldsymbol{B}_q\right) \nonumber\,,
		\end{align}
	\end{subequations}
	with $D_\mu=\partial_\mu + i a^\alpha_{\cl,\mu}\sigma^\alpha\kappa^0+i a^\alpha_{\qn,\mu}\sigma^\alpha\kappa^1$, where the matrices $\kappa^\alpha$ are Pauli matrices acting in cl-q space, with $\kappa^0=\mathbbm{1}_2$.
	At this point, we can write $\boldsymbol{B}=\sqrt{2N}\boldsymbol{b}+\delta\boldsymbol{B}$, with $\sqrt{2N}\boldsymbol{b}=\langle\boldsymbol{B}\rangle$, and integrate out $\delta\boldsymbol{B}$ to get an effective action
	\begin{equation}
		\label{eq: effective action large N}
		\begin{split}
			\mathcal{S}_\mathrm{eff}[\boldsymbol{b},\boldsymbol{A},\boldsymbol{\lambda},\boldsymbol{\vec{H}}]=&\int_{\rr,t}\,\mathcal{L}_\mathrm{HS}[\sqrt{2N}\boldsymbol{b}]\\
			&+2N\mathrm{TrLog}\left(i\mathcal{G}^{-1}[\boldsymbol{A},\boldsymbol{\lambda},\boldsymbol{\vec{H}}]\right)\,,
		\end{split}
	\end{equation}
	where we have defined 
	\begin{equation}
		\begin{split}
			\mathcal{G}^{-1}[\boldsymbol{a},\boldsymbol{\lambda},\boldsymbol{\vec{H}}]=&-D_\mu D^\mu\kappa^1-\gamma(-i\kappa^2)D_t+2i \gamma T(\kappa^0-\kappa^3)\\
			&-\lambda_\cl \,\kappa^1-\lambda_\qn\,\kappa^0-\vec{H}_\cl \cdot\vec{\sigma}\kappa^1-\vec{H}_\qn \cdot\vec{\sigma}\kappa^0
		\end{split}
	\end{equation}
	as a $4\times4$ matrix acting on the twofold gauge index and the $\cl$-$\qn$ one. One can see by inspection in that both terms in Eq.~\eqref{eq: effective action large N} the dependence on $N$ is a simple multiplicative term. Taking the $N\to\infty$ limit makes the saddle point approximation exact, so we can study the equations of motion for the $\boldsymbol{X}=\{\boldsymbol{b},\boldsymbol{a},\boldsymbol{\lambda},\boldsymbol{\vec{H}}\}$ fields:
	\begin{equation}
		\frac{\delta [\mathcal{S}_\mathrm{eff}/(2N)]}{\delta X_\qn}\bigg\rvert_{X_\qn=0}=0\,.
	\end{equation}
	We now assume that the condensates of the fields $X(t)\equiv X_\cl$ depend on time but not on space. We further impose that $\boldsymbol{a}=0$, $\vec{H}_\cl=H(t)\hat{e}_3$, and $b_\cl=(b_1,b_2,0,\dots,0)$. To simplify the equations, we also choose $b_1=b_+(1,0)$ and $b_2=b_-(0,1)$, with $b_\pm\in\mathbb{R}$. We get:
	\begin{subequations}\label{eq:large N eqs1}
		\begin{align}
			& \lambda(t)=r(t)+\frac{u}{2}\left[b_+^2(t)+b_-^2(t)+\int_\qq\left[C^+(q,t)+C^-(q,t)\right]\right],\\
			& H(t) = \frac{v(t)}{2}\left[b_+^2(t)-b_-^2(t)+\int_\qq\left[C^+(q,t)-C^-(q,t)\right]\right],\\
			&\left[\partial_t^2+\gamma\partial_t + \lambda(t)\pm H(t)\right]b_{\pm}(t)=0\,,
		\end{align}
	\end{subequations}
	Where we have defined $C^\pm(q,t)=\left[iG^K_N(|\bq|,t,t)\right]_{\pm,\pm}$, with $G^K_N(\bq,t,t')$ the (matrix) Keldysh ($\cl$-$\cl$) component of the inverse of the operator $\mathcal{G}^{-1}[\boldsymbol{a}=0,\boldsymbol{\lambda}=(\lambda(t),0),\boldsymbol{\vec{H}}=(H(t)\hat{e}_3,0)]$.

	The evolution equations for the large-$N$ Green's function are obtained by inverting the operator $\mathcal{G}^{-1}[\boldsymbol{a}=0,\boldsymbol{\lambda}=(\lambda(t),0),\boldsymbol{\vec{H}}=(H(t)\hat{e}_3,0)]$:
	\begin{widetext}
		\begin{subequations}\label{eq: time evolution Gfs}
			\begin{align}
				&\left[\partial_{t_1}^2 +|\bq|^2 +\gamma\partial_{t_1} + \lambda(t_1)+H(t_1)\sigma^3\right]G^K_N(\bq,t_1,t_2)=4i\gamma T G^A_N(\bq,t_1,t_2)\,,\label{eq: evolution GK}\\
				-&\left[\partial_{t_1}^2 +|\bq|^2 +\gamma\partial_{t_1} + \lambda(t_1)+H(t_1)\sigma^3\right]G^R_N(\bq,t_1,t_2) = \delta(t_1-t_2)\,,\label{eq: evolution GR}\\
				&G^A_N(\bq,t_1,t_2)=\left[G^R_N(\bq,t_2,t_1)\right]^*\,.
			\end{align}
		\end{subequations}
	\end{widetext}
	We now define $\hat{C}(q,t)=i G^K_N(|\bq|,t,t)$, $\hat{D}(q,t)=\partial_{t_1}\partial_{t_2}G^K_N(t_1,t_2)\rvert_{t_1=t_2=t}$ and note that
	\begin{subequations}
		\begin{align}
			&\partial_t \hat{C}(q,t)=2i\partial_{t_1} G^K_N(|\bq|,t_1,t_2)\rvert_{t_1=t_2=t}\,,\\
			&\partial_t^2 \hat{C}(q,t)=2i\partial^2_{t_1} G^K_N(|\bq|,t_1,t_2)\rvert_{t_1=t_2=t} + 2\hat{D}(q,t)\,,\\
			&\partial_t \hat{D}(q,t)=2i\partial^2_{t_1}\partial_{t_2} G^K_N(|\bq|,t_1,t_2)\rvert_{t_1=t_2=t}
		\end{align}
	\end{subequations}
	Setting $t_1=t_2=t$ in Eq.~\eqref{eq: evolution GK}, multiplying it by $i$ on both hand sides, and using the above relations we get
	\begin{multline}
		\label{eq: C eq appendix}
		\frac{1}{2}\partial_t^2\hat{C}(q,t)-\hat{D}(q,t)+\frac{1}{2}\gamma\, \partial_t\hat{C}(q,t)
		\\+[q^2+\lambda(t)+H(t)\sigma^3]\hat{C}(q,t)=-4\gamma T G^A_N(|\bq|,t,t).
	\end{multline}
	Deriving Eq.~\eqref{eq: evolution GK} with respect to $t_2$, multiplying by $i$, and setting $t_1=t_2=t$, we obtain
	\begin{multline}\label{eq: D eq appendix}
		\frac{1}{2}\hat{D}(q,t)+\gamma\hat{D}(q,t)+\frac{1}{2}[q^2+\lambda(t)+H(t)\sigma^3]\partial_t\hat{C}(q,t)\\=-4\gamma T\,\partial_{t_2} G^A_N(|\bq|,t_1,t_2)\rvert_{t_1=t_2=t}\,.
	\end{multline}
	Recalling that 
	\begin{subequations}
		\begin{align}
			&G^A(\bq,t_1,t_2)=\frac{1}{2}\theta(t_2-t_1)\langle[B^\dagger_\bq(t),B_\bq(t')]\rangle \,,\\
			&\partial_{t_2} G^A(\bq,t_1,t_2)=\frac{1}{2}\theta(t_2-t_1)\langle[B^\dagger_\bq(t),\Pi_\bq(t')]\rangle \,,\\
		\end{align}
	\end{subequations}
	with $\theta(0)=1/2$ and $\Pi_q$ the canonically conjugate operator to $B^\dagger_\bq$, and noting that in our theory $[B^\dagger,B]=0$ and $[\Pi,B^\dagger]=1$, we deduce that the right hand sides of Eqs.~\eqref{eq: C eq appendix} and \eqref{eq: D eq appendix} are 0 and $\gamma T$, respectively. Extracting $C^\pm_q(t)=[\hat{C}(q,t)]_{\pm\pm}$ and $D^\pm(q,t)=[\hat{D}(q,t)]_{\pm\pm}$, we get Eqs.~\eqref{eq:EoM2}, \eqref{eq:EoM3}, where we have additionally defined $\lambda_\pm(t)=\lambda(t)\pm H(t)$.
	
	\begin{widetext}
		
		\section{Electromagnetic response}
		\label{sec:em}
		
		In this Appendix, we evaluate the electromagnetic response of some of the phases found for the SU(2) chargon theory in the main text in the large-$N$ limit. 
		\subsection{Electromagnetic response at large-$N$}
		
		At large-$N$, the partition function can be written as 
		\begin{equation}
			\int\mathcal{D}\vec{X} e^{iN\mathcal{S}[\vec{X};A_\mu]}\,,
		\end{equation}
		where $A_\mu$ is an external electromagnetic vector potential that we use to probe the system. For the SU(2) chargon theory we have $\vec{X}=\{\boldsymbol{b},\boldsymbol{a},\boldsymbol{\lambda},\boldsymbol{\vec{H}}\}$ and $\mathcal{S}[\vec{X},A_\mu=0]$ is given by Eq.~\eqref{eq: effective action large N}. The coupling to the electromagnetic field occurs via the covariant derivative $D_\mu=\partial_\mu + i (a^\alpha_{\cl,\mu}\sigma^\alpha+A_{\cl,\mu}\mathbbm{1})\kappa^0+i (a^\alpha_{\qn,\mu}\sigma^\alpha+A_{\qn,\mu}\mathbbm{1})\kappa^1$.
		
		Because we are interested in the linear response, we expand $\mathcal{S}[\vec{X};A_\mu]$ up to second order in $A_\mu$:
		\begin{equation}
			\mathcal{S}[\vec{X};A_\mu]\approx \mathcal{S}_0[\vec{X}]+\left(\mathcal{S}_1^\mu[\vec{X}],A_\mu\right)+\frac{1}{2} \left(A_\mu,\mathcal{S}_2^{\mu\nu}[\vec{X}] A_\nu\right)\,,
		\end{equation}
		where $(\bullet,\bullet)$ is a shorthand for a sum over all internal indices. We now expand the saddle point solution $\vec{X}^*[A_\mu]$ in powers of $A_\mu$:
		\begin{equation}
			\vec{X}^*[A_\mu]=\vec{X}^*_0+\left(\vec{X}_1^{*\mu},A_\mu\right)+\frac{1}{2}\left(A_\mu,\vec{X}_2^{*\mu\nu}A_\nu\right)\,.
		\end{equation}
		The saddle point action is given by 
		\begin{equation}
			\mathcal{S}^*=\mathcal{S}_0[\vec{X}^*_0]+\left(\mathcal{S}^\mu_1[\vec{X}^*_0],A_\mu\right)+\frac{1}{2}\left(A_\mu,\left[\mathcal{S}^{\mu\nu}_2[\vec{X}^*_0]-\Gamma_1^\mu H_0^{-1}\Gamma_1^\nu\right]A_\nu\right)\,,
		\end{equation}
		with $\Gamma_1^\mu=\delta_X \mathcal{S}_1^\mu[\vec{X}^*_0]$ and $H_0=\delta_X^2\mathcal{S}_0[\vec{X}^*_0]$. This tells us that the electromagnetic kernel is given by the tensor $\mathcal{S}^{\mu\nu}_2[\vec{X}^*_0]-\Gamma_1^\mu H_0^{-1}\Gamma_1^\nu$. In particular, we are interested in the spatial components of the $\bq=0$ kernel $K^{\alpha\beta}(t,t')\sim \langle A_{\qn,\alpha}(q=0,t) A_{\cl,\beta}(q=0,t')\rangle$, where $\alpha$ and $\beta$ are only spatial indices. For the special large-$N$ solutions discussed in the main text, the electromagnetic Kernel takes the form:
		\begin{subequations}
			\begin{align}
				K^{\alpha\beta}(t,t')=&K_0^{\alpha\beta}(t,t')-\delta K^{\alpha\beta}(t,t')\,,\\
				K_0^{\alpha\beta}(t,t')=&\left[b^2_1(t)+b^2_2(t)\right]\delta_{t,t'}\delta_{\alpha\beta}+\int_{\bq} \mathrm{Tr}\left[iG^K_N(q;t,t)\right]\delta_{t,t'}\delta_{\alpha\beta}\nonumber\\
				&+i\int_\bq 2q_\alpha q_\beta\mathrm{Tr}\left[G^R_N(q;t,t')G_N^K(q;t',t)+G^K_N(q;t,t')G_N^A(q;t',t)\right] \label{eq: EM Kernel0}\\
				&-\lim_{q\to 0} q_\alpha q_\beta\,\,\mathrm{Tr}\left[G_N^R(q;t,t')\right]\,\left[b_1(t)b_1(t')+b_2(t)b_2(t')\right]\nonumber\\
				&-\lim_{q\to 0} q_\alpha q_\beta\,\,\mathrm{Tr}\left[G_N^R(q;t,t')\sigma^3\right]\,\left[b_1(t)b_1(t')-b_2(t)b_2(t')\right]\nonumber\,,\\
				\delta K^{\alpha\beta}(t,t')=&\int_{t^{\prime\prime},t^{\prime\prime\prime}}\chi_{Aa}^{\alpha\gamma}(t,t^{\prime\prime})D_{aa}^{\gamma\delta}(q=0;t^{\prime\prime},t^{\prime\prime\prime}) \chi_{aA}^{\delta\beta}(t^{\prime\prime\prime},t')\,,\\
				\chi_{Aa}^{\alpha\gamma}(t,t^{\prime\prime})=&[b_1^2(t)-b_2^2(t)]\delta_{t,t'}+\int_\bq \mathrm{Tr}\left[iG_N^K(q;t,t)\sigma^3\right]\delta_{t,t'}\delta_{\alpha\beta}\nonumber\\
				&+i\int_\bq 2q_\alpha q_\gamma\mathrm{Tr}\left[G^R_N(q;t,t')\sigma^3 G_N^K(q;t',t)+G^K_N(q;t,t')\sigma^3 G_N^A(q;t',t)\right]\\
				&-\lim_{q\to 0} q_\alpha q_\beta\,\,\mathrm{Tr}\left[G_N^R(q;t,t')\right]\,\left[b_1(t)b_1(t')-b_2(t)b_2(t')\right]\nonumber\\
				&-\lim_{q\to 0} q_\alpha q_\beta\,\,\mathrm{Tr}\left[G_N^R(q;t,t')\sigma^3\right]\,\left[b_1(t)b_1(t')+b_2(t)b_2(t')\right]\nonumber\,,\\
				\left[D_{aa}^{\alpha\beta}(q=0;t,t')\right]^{-1}=&\left[b^2_1(t)+b^2_2(t)\right]\delta_{t,t'}\delta_{\alpha\beta}+\int_{\bq} \mathrm{Tr}\left[iG^K_N(q;t,t)\right]\delta_{t,t'}\delta_{\alpha\beta}\nonumber\\
				&+i\int_\bq 2q_\alpha q_\beta\mathrm{Tr}\left[\sigma^3G^R_N(q;t,t')\sigma^3G_N^K(q;t',t)+\sigma^3G^K_N(q;t,t')\sigma^3G_N^A(q;t',t)\right]\\
				&-\lim_{q\to 0} q_\alpha q_\beta\,\,\mathrm{Tr}\left[G_N^R(q;t,t')\right]\,\left[b_1(t)b_1(t')+b_2(t)b_2(t')\right]\nonumber\\
				&-\lim_{q\to 0} q_\alpha q_\beta\,\,\mathrm{Tr}\left[G_N^R(q;t,t')\sigma^3\right]\,\left[b_1(t)b_1(t')-b_2(t)b_2(t')\right]\nonumber\,,
			\end{align}
		\end{subequations}
		here $G_N^{R|A|K}(q;t,t')$ are the retarded, advanced and Keldysh components of the Green's function, whose time evolution is governed by equations~\eqref{eq: time evolution Gfs}. Here, $\int_\bq$ is a shorthand for $\int_{q\in[\Lambda_\mathrm{IR},\Lambda_\mathrm{UV}]}\!\frac{d^2\bq}{(2\pi)^2}$. It is interesting to note that in the superconducting state ($H(t)=0$) one finds $\chi_{aA}^{\alpha\beta}=0$, implying $\delta K^{\alpha\beta}=0$. Conversely, in the CDW state ($H(t)\neq0$) $\delta K^{\alpha\beta}$ plays a crucial role in cancelling the Meissner effect coming from $K_0^{\alpha\beta}$. This cancellation is driven by fluctuation of the $a=3$ component of the internal SU(2) gauge field, which hybridizes with the electromagnetic field via $\chi_{aA}^{\alpha\beta}$.
		\subsection{Electromagnetic response in the superconducting steady state}
		We now study the electromagnetic response in a superconducting steady state, that is, where $H(t)=0$ and $b_1(t)=b_2(t)\equiv b(t)$ with $b(t)$ periodic in time. In this case, the Green's function becomes a scalar and we can represent it using Floquet theory as an infinite matrix.
		\begin{subequations}
			\begin{align}
				&\left[\boldsymbol{G}^{R|A}(q,\omega)\right]^{-1}_{nm}=\left[(\omega+n\Omega)^2+i\gamma(\omega+n\Omega)-q^2\right]\delta_{nm}-\sum_{\ell=-\infty}^{+\infty}\lambda_\ell\, \delta_{n,m+\ell}\,,\\
				&\boldsymbol{G}^K(q,\omega)=-2i\gamma \boldsymbol{G}^R(q,\omega) \boldsymbol{F}(\omega) \boldsymbol{G}^A(q,\omega)\,,\\
				&[\boldsymbol{F}(\omega)]_{mm}=(\omega+n\Omega) n_B(\omega+n\Omega)\,\delta_{mn}\,,
			\end{align}
		\end{subequations}
		with $n_B(x)=1/(e^{x/T}-1)$ the Bose-Einstein distribution function. In the classical (high-temperature) limit considered in the main text, the distribution function becomes $\boldsymbol{F}(\omega)=2T\delta_{mn}$. For simplicity, we work in the Coulomb gauge, $\vec{\nabla}\cdot \vec{A}=0$, which allows us to neglect the last two terms in Eq.~\eqref{eq: EM Kernel0}. In general, those terms will ensure that the electromagnetic kernel stays gauge invariant. In the steady state, we can expand the kernel in Floquet harmonics as
		\begin{equation}
			K^{\alpha\beta}(t,t')=\int_{-\infty}^{+\infty}\!\frac{d\omega}{2\pi}\,e^{-i\left(\omega-\frac{n}{2}\Omega\right)(t-t')}\sum_{n=-\infty}^{+\infty}e^{-in\Omega\frac{t+t'}{2}}\,\mathcal{K}^{\alpha\beta}_n(\omega)\,.
		\end{equation}
		In the following, we will focus only on the period-averaged (or homodyne) electromagnetic kernel $\mathcal{K}^{\alpha\beta}_0(\omega)$. It reads
		\begin{equation}
			\begin{split}
				\mathcal{K}_0^{\alpha\beta}(\omega)=2\sum_{n=-\infty}^{+\infty}b_n b_{-n}\delta_{\alpha\beta}+&i\int_\bq 4q_\alpha q_\beta\int_{-\infty}^{+\infty}\!\frac{d\nu}{2\pi}\,\sum_{n=-\infty}^{+\infty}\left(\boldsymbol{G}^R_{0,n}(\nu)\boldsymbol{G}^K_{n,0}(\nu-\omega)+\boldsymbol{G}^K_{0,n}(\nu)\boldsymbol{G}^A_{n,0}(\nu-\omega)\right)\\
				&+\int_\bq\int_{-\infty}^{+\infty}\!\frac{d\nu}{2\pi}\,2i\boldsymbol{G}^K_{0,0}(\nu)\,\delta_{\alpha\beta}\,.
			\end{split}
		\end{equation}
		Integrating the last term by parts, we can see that the (period-averaged) Meissner effect is given by the simple formula $\mathcal{K}_0^{\alpha\beta}(\omega\to0)=2\sum_{n=-\infty}^{+\infty}b_n b_{-n}\,\delta_{\alpha\beta}$. In the general case where $b_1(t)$ and $b_2(t)$ are unequal and $H(t)$ nonzero, one can show that the cancellation between the "paramagnetic" and "diamagnetic" terms occurs also for $\chi_{aA}^{\alpha\beta}$ and $D_{aa}^{\alpha\beta}$ in the steady state, so that the Meissner effect is given by
		\begin{equation}
			\int_{-\infty}^{+\infty}\!d\tau\,K^{\alpha\beta}(t+\tau/2,t-\tau/2) = \left\{b_1^2(t)+b_2^2(t) - \frac{[b_1^2(t)-b_2^2(t)]^2}{b_1^2(t)+b_2^2(t)}\right\}\delta_{\alpha\beta}\,.
		\end{equation}
		This shows that the CDW phase $(b_2(t)=0)$ has no Meissner effect, as expected.
		
		The optical conductivity can be obtained from the kernel as 
		\begin{equation}
			\sigma^{\alpha\beta}_n(\omega)=\frac{\mathcal{K}_n^{\alpha\beta}(\omega)}{i\omega}\,.
		\end{equation}
	\end{widetext}

\end{document}